\documentclass[runningheads]{llncs}
\usepackage[T1]{fontenc}
\usepackage{graphicx}
\usepackage{amsmath, amssymb}
\usepackage{bm}

\DeclareMathOperator{\sgn}{sgn}

\begin{document}

\title{Diffusion--Shock Inpainting
\thanks{This project has received funding from the European Research Council
(ERC) under the European Union’s Horizon 2020 research and innovation
programme (grant agreement No. 741215, ERC Advanced Grant INCOVID).}}
\titlerunning{DS Inpainting}
%
\author{Kristina Schaefer\and Joachim Weickert}
\authorrunning{K. Schaefer and J. Weickert}

%
\institute{Mathematical Image Analysis Group, 
Dept.~of Mathematics and Computer Science, E1.7,
Saarland University,
66123 Saarbr\"ucken, Germany\\
\email{$\{$schaefer, weickert$\}$@mia.uni-saarland.de}}

\maketitle              

\begin{abstract}
We propose diffusion--shock (DS) inpainting as a hitherto unexplored 
integrodifferential equation for filling in missing structures 
in images. It combines two carefully chosen components that 
have proven their usefulness in different applications: homogeneous 
diffusion inpainting and coherence-enhancing shock filtering. 
DS inpainting enjoys the complementary synergy of its 
building blocks: It offers a high degree of anisotropy along an 
eigendirection of the structure tensor. This enables it to 
connect interrupted structures over large distances. Moreover,
it benefits from the sharp edge structure generated by the shock 
filter, and it exploits the efficient filling-in effect of homogeneous 
diffusion. The second order equation that underlies DS inpainting
inherits a continuous maximum--minimum principle from its constituents. 
In contrast to other attractive second order inpainting equations such as 
edge-enhancing anisotropic diffusion, we can guarantee this property also 
for the proposed discrete algorithm. Our experiments show a performance 
that is comparable to or better than many linear or nonlinear, isotropic 
or anisotropic processes of second or fourth order. They include
homogeneous diffusion, biharmonic interpolation, TV inpainting, 
edge-enhancing anisotropic diffusion, the methods of Tschumperl\'e and 
of Bornemann and M\"arz, Cahn--Hilliard inpainting, and Euler's elastica. 

\end{abstract}


\section{Introduction}

Image inpainting~\cite{EL99a,MM98a} aims at restoring images 
with missing or damaged areas. Many popular methods use partial 
differential equations (PDEs), since they offer compact and transparent 
models with sound theoretical foundations. A particularly simple representative
is homogeneous diffusion inpainting~\cite{Ca88}. It is based on a
linear second order PDE that satisfies a maximum--minimum principle
and allows to design very efficient algorithms \cite{KW22}. For 
inpainting-based compression, it may give very good results, if the data 
are optimised carefully~\cite{MHWT12}. However, it cannot produce 
satisfactory continuations of sharp edges over large distances. 

\smallskip
As a rememdy, higher order PDEs have been considered, such as the
ones arising from the energy functionals of Euler's elastica 
\cite{KTZ19,MM98a,Mu94a} or the Cahn--Hilliard model \cite{BEG07,BHS09}. 
From a theoretical perspective, these continuous models are attractive: 
They offer a low-curved continuation of level lines or the propagation 
of gradient information. Unfortunately, for such higher order PDEs it 
is fairly challenging to design good numerical algorithms that are 
computationally efficient and do not suffer from dissipative artefacts, 
which lead to a blurred continuation of edges. 

\smallskip
An interesting alternative to higher order inpainting PDEs are 
second order anisotropic integrodifferential methods that implicitly 
exploit curvature information via a Gaussian convolution inside a 
diffusion tensor. This idea is pursued in edge-enhancing anisotropic 
diffusion (EED) \cite{WW06}. It achieves state-of-the-art results for
inpainting-based compression and can propagate structures over large 
distances \cite{SPME14}. However, current discretisations can violate 
a maximum--minimum principle. Moreover, although edges remain sharper 
than for most algorithms for higher order PDEs, they still show 
some dissipative artefacts. 
 
\begin{figure}[tb]
\centering
\setlength{\fboxsep}{1pt}
\begin{tabular}{c@{\hspace{2mm}}c}
\fbox{\includegraphics[width=0.3\textwidth]{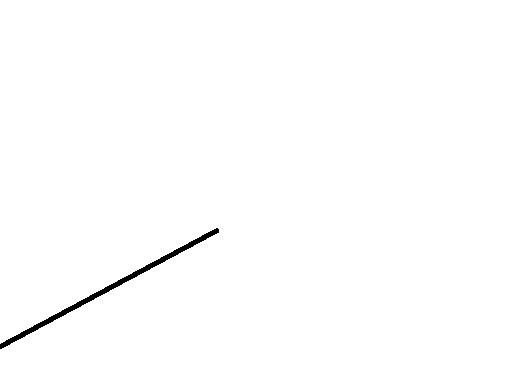}} &
\fbox{\includegraphics[width=0.3\textwidth]{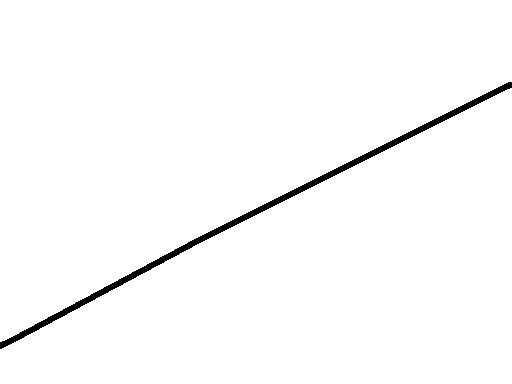}}\\
original & steady state
\end{tabular}
\caption{Line elongation with the coherence-enhancing shock filter. 
\textbf{Left}: Original image, $512\times 384$. \textbf{Right}: 
Steady state of the shock filter, $\sigma= 2$, $\rho = 5$. }
\label{fig:line}
\end{figure}

\medskip
\textbf{Our Contribution.}
The goal of our work is to show that all the above mentioned problems
can be addressed with a surprisingly simple combination of two processes 
with complementary qualities: homogeneous diffusion inpainting~\cite{Ca88} 
and a coherence-enhancing shock filter~\cite{We03}. While both techniques  
are well-established, their combination within an inpainting method is 
novel. The resulting {\em diffusion--shock (DS) inpainting} offers the
best of two worlds: On the one hand, the coherence-enhancing shock filter 
is guided by the  robust edge information from a structure tensor 
\cite{FG87}. Its switch between the hyperbolic PDEs of dilation and 
erosion creates perfectly sharp shock fronts that can be propagated 
with basically no directional artefacts. This is illustrated in
Fig.~\ref{fig:line}. On the other hand, the parabolic homogeneous 
diffusion PDE is a model with maximal simplicity that creates 
an efficient filling-in mechanism in flat areas. For both processes 
we use discretisations that offer a high degree of rotation invariance
and satisfy a discrete maximum--minimum principle. These properties
carry over to DS inpainting. In our experiments we compare DS 
inpainting  with many inpainting PDEs and demonstrate its favourable
performance in spite of its simplicity.
 

\medskip
\textbf{Related Work.}
With the goal of image sharpening, Kramer and Bruckner~\cite{KB75} proposed the 
first discrete definition of a shock filter, before Osher and Rudin gave the 
first PDE-based formulation~\cite{OR90}.  In both cases, the morphological 
operations dilation and erosion with a disk-shaped structuring element are 
applied adaptively, depending on the sign of a second derivative
operator. To make the process more robust against noise, Alvarez and Mazorra 
proposed to presmooth the image before computing this second derivative 
operator that guides the process~\cite{AM94}. 
Weickert's coherence-enhancing shock filter \cite{We03} uses the second 
derivative in direction of the dominant eigenvector of the structure tensor. 
Welk et al.~\cite{WWG07} proved well-posedness results for 1-D semidiscrete 
and discrete shock filters. 
It is well-known that several classical image enhancement methods 
implicitly or explicitly combine a smoothing PDE with a shock filter; 
see e.g.~\cite{KDA97,Bo02}. Such combinations, however, rarely appear 
in inpainting applications.  
 
\smallskip
Apart from the already mentioned PDE-based inpainting methods, there
some additional ones that play a role in our performance evaluation.
Biharmonic interpolation \cite{Du76} is the fourth order counterpart to 
homogeneous diffusion inpainting and offers reasonable quality in 
inpainting-based compression \cite{SPME14}. Total variation (TV) inpainting 
\cite{SC02} can be seen as a limit case of Perona--Malik \cite{PM90} 
inpainting with a Charbonnier diffusivity \cite{CBAB97}. Tschumperl\'e's 
method \cite{Ts06} involves a tensor-driven equation that uses the 
curvature of integral curves. This also qualifies it as candidate for
bridging interrupted structures over large distances.

\smallskip
While many inpainting PDEs are elliptic or parabolic, hyperbolic ones 
such as the dilations and erosions in shock filters are rarely used 
within inpainting methods.  
A recent inpainting model by Novak and Reini\'c \cite{NR22} combines 
a shock filter with the fourth order Cahn--Hilliard PDE. Our DS 
inpaiting is conceptually simpler: already a second order homogeneous 
diffusion PDE suffices to achieve the desired filling-in effect.
The method of Bornemann and M\"arz \cite{BM07} is closest in spirit 
to DS inpainting. It uses transport processes that are guided by a 
structure tensor. In contrast to our work, however, their paper follows 
a procedural--algorithmic approach without specifying a compact 
integrodifferential equation. In our experiments we will compare 
against this method. 


\medskip
\textbf{Paper Structure.} In Section \ref{sec:shock-filter}, we review the 
concept of coherence-enhan\-cing shock filtering since it is fundamental for 
our work. Section \ref{sec:inpainting} introduces our proposed DS inpainting
model. A corresponding algorithm is discussed in Section~\ref{sec:algorithm},
followed by an experimental evaluation in Section \ref{sec:experiments}. 
Section~\ref{sec:conclusion} concludes the paper and gives an outlook to 
future work. 


\section{Review of Coherence-Enhancing Shock Filtering}
\label{sec:shock-filter}
Since DS inpainting substantially relies on coherence-enhancing shock
filters and their PDE-based definitions of dilation and erosion, we
first review these concepts, in order to keep our paper self-contained. 


\subsection{PDE-based Morphology}
Let $f: \Omega \to\mathbb{R}$ denote a greyscale image on a rectangular
image domain $\Omega\subset\mathbb{R}^2$. Mathematical morphology \cite{So04}
considers a neighbourhood (structuring element) $B$. The dilation 
$\delta_B[f]$ replaces $f$ by its supremum within $B$, and the erosion 
$\varepsilon_B[f]$ uses an infimum instead: 
\begin{align}
\delta_B[f](\bm x ) \; &= \; 
    \sup\{ f(\bm x -\bm y)\; \mid \; \bm y \in B \}\,,\\
\varepsilon_B[f](\bm x ) \; &= \; 
     \inf\; \{ f(\bm x+\bm y)\; \mid \; \bm y \in B \}\,.
\end{align}
If $B$ is a disk of radius $t$, dilation and erosion $u(\bm x, t)$ follow the 
PDE \cite{BM92}
\begin{align}
\partial_t u \; = \; \pm  \lvert\bm \nabla u\rvert 
\end{align}
with initial image $u(\bm x, 0) = f(\bm x)$ and reflecting boundaries.
The $+$ sign describes dilation, and the $-$ sign erosion. 
By $\bm \nabla = (\partial_x, \partial_y)^\top$ we denote the spatial 
nabla operator, and $|\cdot|$ is the Euclidean norm.  


\subsection{Coherence-Enhancing Shock Filtering}
\label{sec:shock}
The dilation PDE propagates the grey values of local maxima, while erosion 
propagates the grey values of local minima. To enhance the sharpness of 
images, shock filters locally apply dilation in influence zones of maxima and 
erosion in influence zones of minima. The sign of a second order derivative 
operator determines these influence zones. Shocks are formed at its zero
crossings. Individual shock filters differ in their second order 
derivative operator, which may also involve some smoothing to make 
the filter more robust under noise. 

\smallskip
The coherence-enhancing shock filter \cite{We03} that we use is 
governed by the PDE
\begin{align}
\partial_t u \; = \;
         - \sgn\big(\partial_{\bm w\bm w}u_\sigma\big) |\bm \nabla u|
\end{align}
with initial image $u(\bm x, 0) = f(\bm x)$ and reflecting boundary
conditions. It involves the second derivative along the dominant 
eigenvector $\bm w$ (with the larger eigenvalue) of the structure tensor 
\cite{FG87}. Let $u_\sigma = K_\sigma \ast u$ denote the convolution of 
$u$ with a Gaussian $K_\sigma$ of standard deviation $\sigma$.
Then the structure tensor is given by a componentwise convolution 
of the matrix $\bm{\nabla} u_\sigma \, \bm{\nabla} u_\sigma^\top$ 
with a Gaussian $K_\rho$:
\begin{equation}
\bm J_\rho(\bm \nabla u_\sigma) \;=\;
  K_\rho * \left(\bm{\nabla} u_\sigma \, \bm{\nabla} u_\sigma^\top\right).
\end{equation}
Its eigenvalues describe the average quadratic contrast in the direction
of the eigenvectors. Thus, steering the shock filter with 
$\partial_{\bm w\bm w}u_\sigma$ encourages the formation of coherent,
flow-like structures with maximal contrast in direction $\bm{w}$.  
This justifies the name of the coherence-enhancing shock filter.
For $t\to \infty$ it leads to a typically non-flat steady state 
that shows elongated image structures with very sharp edges.
In contrast to \cite{We03}, we use $\bm J_\rho(\bm \nabla u_\sigma)$ 
instead of $\bm J_\rho(\bm \nabla u)$, which yields better results 
in our case.

\smallskip
Fig.~\ref{fig:line} illustrates the performance of this shock filter. 
Without visible deviations, it elongates the black line over more
than 200 pixels in a direction that is not grid aligned. Moreover, the 
result is extremely sharp without any dissipative artefacts. This quality 
is hardly ever seen in PDE-based algorithms. 


\section{Diffusion--Shock Inpainting}
\label{sec:inpainting}

Coherence-enhancing shock filtering is designed to propagate information
only in coherence direction. The scale of the created structures perpendicular 
to its propagation direction is determined by the presmoothing scale $\sigma$. 
Hence, it cannot fill in large homogeneous areas beyond that scale. 
Here, homogeneous diffusion inpainting \cite{Ca88} may serve as an ideal 
partner. It is simple, parameter-free, satisfies a maximum--minimum 
principle, and offers an efficient isotropic filling-in mechanism 
by solving $\,0 = \Delta u = \partial_{xx}u + \partial_{yy}u\,$ in 
inpainting regions.

\smallskip
Therefore, we design our {\em diffusion--shock (DS) inpainting} such 
that it performs a convex combination of both processes: Around edges,
the coherence-enhancing shock filter from Subsection \ref{sec:shock}
dominates, and homogeneous diffusion inpainting is activated in flat 
regions. Let the locations of the known values be given by a so-called 
{\em inpainting mask} $K\subset \Omega$. Here we do not alter the grey 
values. In the inpainting domain $\Omega \setminus K$, the reconstruction 
is computed as the steady state ($t\to\infty$) of the integrodifferential 
equation
\begin{equation}
\label{eq:inpainting}
\partial_t u \; = \; g\left(|\bm \nabla u_\nu|^2\right)\, \Delta u
  \;-\; \Big(1-g\left(| \bm \nabla u_\nu |^2\right)\!\Big) \,
  \sgn\big( \partial_{\bm w\bm w} (u_\sigma) \big) \,
  | \bm \nabla u | 
\end{equation}
with Dirichlet data at the boundaries $\partial K$ of the inpainting mask,
and reflecting boundary conditions on the image domain boundary $\partial 
\Omega$. By $u_\nu=K_\nu \ast u$ we denote a convolution of $u$ with 
a Gaussian of standard deviation $\nu$. The weight function
$g\!:\mathbb{R}\to\mathbb{R}$ is a nonnegative decreasing function with
$g(0)=1$, and $g(s^2) \to 0$ for $s^2\to\infty$.
It has the same structure as a diffusivity function in nonlinear diffusion 
filters \cite{PM90}. We use the Charbonnier diffusivity \cite{CBAB97}
\begin{equation}
  g\left(s^2\right) \; = \; \frac{1}{\sqrt{1+ s^2/\lambda^2}}
\end{equation} 
with a contrast parameter $\lambda > 0$. Thus, the weight 
$g\left(|\bm \nabla u_\nu|^2\right)$ secures a smooth transition between
the shock term and the homogeneous diffusion term. Its Gaussian scale 
$\nu$ makes it robust under noise. 


\section{Numerical Algorithm}
\label{sec:algorithm}
Continuous DS inpainting inherits a maximum--minimum principle from its 
diffusive and morphological components. To obtain an algorithm for DS 
inpainting that preserves this property, we discretise equation 
(\ref{eq:inpainting}). We assume the same grid size $h > 0$ in $x$- 
and $y$-direction. Let $\tau>0$ denote the time step size, and let 
$u_{i,j}^k$ be an equal discrete approximation of $u(\bm{x},t)$ in pixel 
$(i,j)$ at time $k\tau$.
For simplicity, we use an explicit finite difference scheme. It
approximates $\partial_t u$ at time level $k$ with the forward difference
\begin{equation}
 \label{eq:forward}
 (\partial_t u)_{i,j}^k \;=\; \frac{u_{i,j}^{k+1}-u_{i,j}^k}{\tau}
\end{equation}
and evaluates the right hand side of (\ref{eq:inpainting}) at the old time
level $k$. Thus, let us now focus on the space discretisation of the
diffusion term and the shock term. 

\smallskip
For the homogeneous diffusion term $\Delta u$ classical central differences 
are suitable. Welk and Weickert show in \cite{WW21} that a weighted 
combination of axial and diagonal central differences with weight 
$\delta = \sqrt{2}-1 $ results in a scheme with good rotation invariance. 
The corresponding stencil is given by 
\begin{equation}
\mbox{
$(\Delta u)_{i,j}^k$ $\;=\;$ 
$\Biggl( \dfrac{1-\delta}{h^2}$
\begin{tabular}{|c|c|c|}
\hline
$\;0$ & $\;1$ & $\;0$ \\
\hline
$\;1$ & $-4$ & $\;1$ \\
\hline
$0$ & $1$ & $0$ \\
\hline
\end{tabular}
$ \;+\;  \dfrac{\delta}{2h^2}$
\begin{tabular}{|c|c|c|}
\hline
$\;1$ & $\;0$ & $\;1$ \\
\hline
$\;0$ & $-4$ & $\;0$ \\
\hline
$\;1$ & $\;0$ & $\;1$ \\
\hline
\end{tabular}
$\Biggr) \, u_{i,j}^k\,.$
}
\end{equation}
With this stencil in space and the forward 
difference (\ref{eq:forward}) in time we obtain an explicit scheme for
the homogeneous diffusion equation $\partial_t u = \Delta u$. If
\begin{equation}
 \label{eq:taud}
 \tau \;\leq\; \frac{h^2}{4-2\delta} \;=:\; \tau_D\,,
\end{equation}
one easily verifies that $u_{i,j}^{k+1}$ is a convex combination 
of data at time level $k$. This implies stability in terms of the 
maximum--minimum principle
\begin{align}
\label{eq:maxmin}
\min_{n,m} f_{n,m} \; \leq\; u^k_{i,j} \;\leq\; \max_{n,m} f_{n,m}\; 
\qquad  \mbox{for all $i$, $j$, and for $k \geq 1$.} 
\end{align}

A space discretisation of morphological evolutions of type
$\partial_t u = \pm |\bm \nabla u|$ is not as straightforward.
Typically one uses upwind methods such as the 
Rouy--Tourin scheme~\cite{RT92}. In order to improve its rotation 
invariance, we follow Welk and Weickert \cite{WW21} again, who propose a 
weighted combination of a Rouy--Tourin scheme in axial and diagonal 
direction. For the dilation term $|\bm \nabla u|$ this yields
\begin{align}
|\bm \nabla u|_{i,j}^k \;=\;
\tfrac{1-\delta}{h} \: \big(
       &\max \, \lbrace u_{i+1,j}^k \!-\! u_{i, j}^k,\;
                     u_{i-1,j}^k \!-\! u_{i, j}^k,\; 0\rbrace^2
                        \nonumber\\
     + &\max \, \lbrace u_{i,j+1}^k \!-\! u_{i, j}^k,\; 
                     u_{i,j-1}^k \!-\! u_{i, j}^k,\;0\rbrace^2
     \big)^\frac{1}{2} \nonumber\\[1mm]
+\;\tfrac{\delta}{\sqrt{2}h} \: \big(
       &\max \, \lbrace u_{i+1,j+1}^k \!-\! u_{i, j}^k,\;
                        u_{i-1,j-1}^k \!-\! u_{i, j}^k,\; 0\rbrace^2
                        \nonumber\\
     + &\max \, \lbrace u_{i-1,j+1}^k \!-\! u_{i, j}^k,\; 
                     u_{i+1,j-1}^k \!-\! u_{i, j}^k,\;0\rbrace^2
      \big)^\frac{1}{2}
\label{eq:dis-dil}     
\end{align}
with some weight $\delta \in [0,1]$.
For the erosion term $-|\bm \nabla u|$, we use
\begin{align}
-|\bm \nabla u|_{i,j}^k \;=\;
-\tfrac{1-\delta}{h} \: \big(
       &\max \, \lbrace u_{i,j}^k \!-\! u_{i+1, j}^k,\;
                        u_{i,j}^k \!-\! u_{i-1, j}^k,\; 0\rbrace^2
                        \nonumber\\
     + &\max \, \lbrace u_{i,j}^k \!-\! u_{i, j+1}^k ,\; 
                         u_{i,j}^k \!-\! u_{i, j-1}^k,\;0\rbrace^2
     \big)^\frac{1}{2} \nonumber\\[1mm]
-\;\tfrac{\delta}{\sqrt{2}h} \: \big(
       &\max \, \lbrace u_{i,j}^k \!-\! u_{i+1,j+1}^k,\;
                        u_{i,j}^k \!-\! u_{i-1,j-1}^k,\;0\rbrace^2
                        \nonumber\\
     + &\max \, \lbrace u_{i,j}^k \!-\! u_{i-1,j+1}^k ,\; 
                         u_{i,j}^k \!-\! u_{i+1,j-1}^k ,\;0 \rbrace^2
      \big)^\frac{1}{2}         
      \;.
\label{eq:dis-ero}  
\end{align}
It can be shown that an explicit scheme with time discretisation 
(\ref{eq:forward}) and space discretisation \eqref{eq:dis-dil} or 
\eqref{eq:dis-ero} satisfies the maximum--minimum principle~(\ref{eq:maxmin}) 
if
\begin{equation}
 \label{eq:taum}
 \tau \;\leq\; \frac{h}{\sqrt{2}\,(1-\delta) + \delta} \;=:\; \tau_M\,.
\end{equation}
For approximating rotation invariance well, we follow \cite{WW21} and
set $\delta =\sqrt{2} - 1$.

\smallskip
To go from dilation and erosion to a coherence-enhancing shock filter,
we have to discretise $\partial_{\bm w  \bm w} u_\sigma$. 
We compute all Gaussian convolutions in the spatial domain with a
sampled and renormalised Gaussian. To guarantee a high approximation
quality, it is truncated at five times its standard deviation.                 
In order to implement reflecting boundary conditions, we add an
extra layer of mirrored dummy pixels around the image domain.
For the first order derivatives within the structure tensor,
we enforce this mirror symmetry by imposing zero values at
the image boundaries.
We approximate $\partial_x u$ and $\partial_y u$ in the structure 
tensor by means of Sobel operators \cite{DH73}, which offer a high
degree of rotation invariance. Since the structure tensor is a 
symmetric $2 \times 2$ matrix, its normalised dominant eigenvector 
$\bm w = (c, s)^\top$ can be computed analytically. Moreover, we 
have 
\begin{equation}
 \left(\partial_{\bm w  \bm w} v\right)_{i,j} ^k
 \;=\; \left(c^2 \, \partial_{xx}v + 2 cs \, \partial_{xy}v + 
                   s^2 \, \partial_{yy}v\right)_{i,j}^k.
\end{equation} 
All second order derivatives on the right hand side are approximated 
with their standard central differences.  

\smallskip
Putting everything together yields the following explicit scheme for 
\eqref{eq:inpainting}:
\begin{align}
\frac{u^{k+1}_{i,j}- u^k_{i,j}}{\tau}\;=\;
  g_{i,j}^k \cdot \big(\Delta u\big)_{i,j}^k - (1-g_{i,j}^k)\cdot 
  \sgn\big((\partial_{\bm w  \bm w} u_\sigma)^k_{i,j}\big) 
  \, |\bm \nabla u|^k_{i,j}
\label{eq:discrete}
\end{align}
with initial condition $u^0_{i,j} = f_{i,j}$. It inherits its stability
from the stability results of the schemes for diffusion and morphology:


\begin{theorem} {\bf (Stability of the DS Inpainting Scheme)}\\[1mm]
Let the time step size $\tau$ of the scheme \eqref{eq:discrete}
be restricted by 
\begin{equation}
\tau \; \leq \;  \min\,\{\tau_D, \, \tau_M\}
\end{equation}
with $\tau_D$ and $\tau_M$ as in (\ref{eq:taud}) and (\ref{eq:taum}).\\[1mm]
Then the scheme satisfies the discrete maximum--minimum principle
\begin{align}
\min_{n,m} f_{n,m} \; \leq\; u^k_{i,j} \;\leq\; \max_{n,m} f_{n,m}\;
\qquad  \mbox{for all $i$, $j$, and for $k \geq 1$.}
\end{align}
\end{theorem}

\smallskip
\begin{proof}
If $\tau \leq \min\,\{\tau_M, \, \tau_D\}$, it follows from the stability of 
the diffusion and morphological processes that
\begin{align*}
u^{k+1}_{i,j} \;&=\; u^k_{i,j} +\tau g_{i,j}^k \cdot \big(\Delta u\big)_{i,j}^k
 - (1-g_{i,j}^k)\cdot\,\tau \sgn\big((\partial_{\bm w  \bm w}u_\sigma)^k_{i,j}\big) 
  \, |\bm \nabla u|^k_{i,j}\\
&\leq\; g_{i,j}^k \max_{n,m} f_{n,m} + (1-g_{i,j}^k) 
 \max_{n,m} f_{n,m} \\
 \;&=\;\max_{n,m} f_{n,m} \; .
\end{align*}
Analogously, one can show the condition 
$\, \min\limits_{n,m} f_{n,m} \, \leq \, u^k_{i,j}$. \qed
\end{proof}
Thus, for $\,\delta = \sqrt{2}-1\,$ and $\,h=1\,$ our 
scheme is stable for $\,\tau \,\leq\, \tau_D \,\approx\, 0.31$. 
Theorem 1 shows an advantage of DS inpainting over EED inpainting 
\cite{SPME14,WW06}, for which a discrete maximum--minimum principle cannot 
be guaranteed so far. 


\section{Experiments}
\label{sec:experiments}

Let us now evaluate DS inpainting experimentally. We mainly focus on 
binary images, since their high contrast is especially vulnerable to 
dissipative artefacts, but also present one experiment with greyscale
images. Whenever the mask image is given, the white area denotes the 
inpainting mask, and the black area depicts the unknown regions. All 
methods in our evaluation use optimised parameters. 

\begin{figure}[tbp]
\centering
\setlength{\fboxsep}{1pt}

\begin{tabular}{c@{\hspace{2mm}}c@{\hspace{2mm}}c}
\fbox{\includegraphics[width=0.25\textwidth]{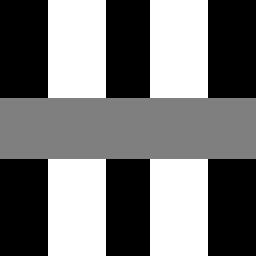}} &
\fbox{\includegraphics[width=0.25\textwidth]{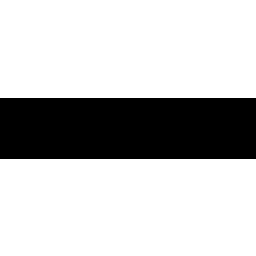}} &
\fbox{\includegraphics[width=0.25\textwidth]{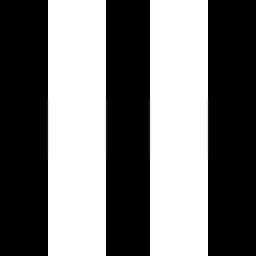}} \\[2mm] 
\fbox{\includegraphics[width=0.25\textwidth]{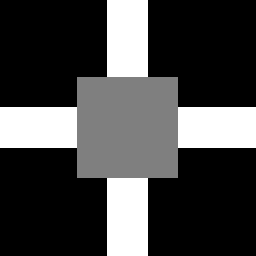}} &
\fbox{\includegraphics[width=0.25\textwidth]{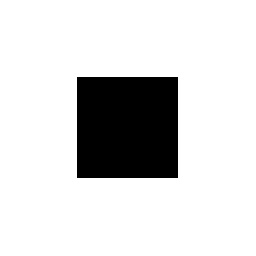}} &
\fbox{\includegraphics[width=0.25\textwidth]{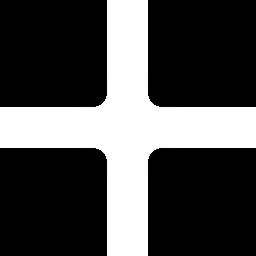}}\\
input & mask & DS inpainting
\end{tabular}
\caption{DS inpainting of $256\times256$ test images. Parameters: 
 \textbf{Top}: $\sigma = 2$, $\rho=5$, $\nu=3$, and $\lambda = 3$.
 \textbf{Bottom}:
 $\sigma = 2$, $\rho=5$, $\nu=2$, and $\lambda = 2$.}
\label{fig:cahn}
\end{figure}

\smallskip
In Fig. \ref{fig:cahn}, we apply DS inpainting to shape completion problems 
inspired by the experiments for Cahn--Hilliard inpainting in \cite{BEG07}. 
Our operator connects bars and restores a cross while maintaining the 
high contrast of the binary images. Compared to the results in \cite{BEG07}, 
DS inpainting offers sharper reconstructions. 

\begin{figure}[tbp]
\centering
\setlength{\fboxsep}{1pt}
\begin{tabular}{c@{\hspace{2mm}}c@{\hspace{2mm}}c}
\fbox{\includegraphics[width=0.25\textwidth]{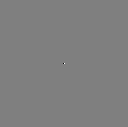}} &
\fbox{\includegraphics[width=0.25\textwidth]{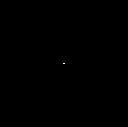}} &
\fbox{\includegraphics[width=0.25\textwidth]{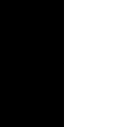}} \\[2mm]
\fbox{\includegraphics[width=0.25\textwidth]{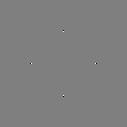}} &
\fbox{\includegraphics[width=0.25\textwidth]{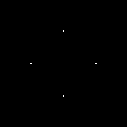}} &
\fbox{\includegraphics[width=0.25\textwidth]{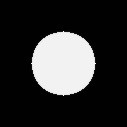}}\\
input  & mask & DS inpainting 
\end{tabular}
\caption{DS inpainting from dipoles.
\textbf{Top}: $128\times 128$ image; $\sigma = 1$, $\rho=2$, $\nu=2$, 
 $\lambda = 1$.
\textbf{Bottom}:  $127\times 127$ image;
$\sigma = 2.65$, $\rho=4$, $\nu=2$, $\lambda = 3$.}
\label{fig:dipoles}
\end{figure}

\smallskip
Fig.~\ref{fig:dipoles} shows a more challenging experiment, inspired by
\cite{SPME14}. It drives the sparsity of the inpainting data to the extreme
by specifying only one or four dipoles. Nevertheless, DS inpainting shows
a flawless performance: It creates two sharp half planes from one dipole, 
and a disk from four dipoles. This demonstrates its ability to bridge large
distances and its curvature reducing effect.

\begin{figure}[tbp]
\centering
\setlength{\fboxsep}{1pt}
\begin{tabular}{c@{\hspace{2mm}}c@{\hspace{2mm}}c@{\hspace{2mm}}c}
input &  hom. diff. & biharmonic &  TV \\
{\includegraphics[width=0.23\textwidth]{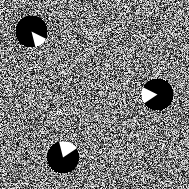}} &
{\includegraphics[width=0.23\textwidth]{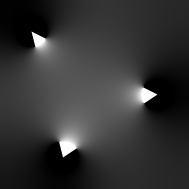}} &
{\includegraphics[width=0.23\textwidth]{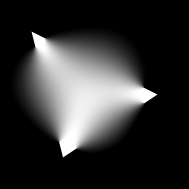}} &
{\includegraphics[width=0.23\textwidth]{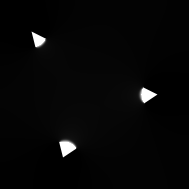}} \\[1mm]
{\includegraphics[width=0.23\textwidth]{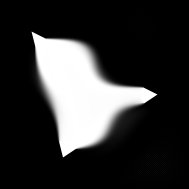}}&
{\includegraphics[width=0.23\textwidth]{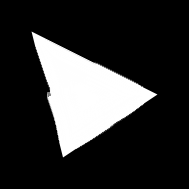}} &
{\includegraphics[width=0.23\textwidth]{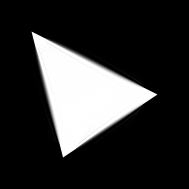}} &
{\includegraphics[width=0.23\textwidth]{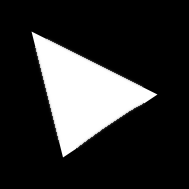}}\\
Tschumperl\'e &BM & EED & DS 
\end{tabular}
\caption{Comparison of inpainting methods. \textbf{Top}:
 Input image with known data in the disks and noise in the unknown region,
 homogeneous diffusion, biharmonic interpolation, and TV inpainting.
\textbf{Bottom}: Tschumperl\'e's approach, Bornemann--M\"arz (BM) method,
 EED inpainting, DS inpainting with $\sigma = 4.7$, $\rho=6$, $\nu=5.2$, and
 $\lambda = 3.4$.}
\label{fig:kani}
\end{figure}

\smallskip
Fig.~\ref{fig:kani} shows an exhaustive comparison of DS inpainting
with many results from Schmaltz et al.~\cite{SPME14}. The goal is
to reconstruct a Kaniza-like triangle. Homogeneous 
diffusion \cite{Ca88}, biharmonic interpolation~\cite{Du76},
and TV inpainting \cite{SC02} are unable to produce sharp edges. 
Tschumperl\'e's algorithm \cite{Ts06} connects the corners, but fails to 
produce correctly oriented sharp edges. The Bornemann--M\"arz approach
\cite{BM07}, edge-enhancing anisotropic diffusion (EED) \cite{SPME14,WW06}, 
and DS inpainting reconstruct a satisfactory triangle. DS inpainting 
offers the best overall quality with high directional
accuracy and no visible dissipative artefacts. 

\begin{figure}[tb]
\centering
\setlength{\fboxsep}{1pt}
\begin{tabular}
{c@{\hspace{2mm}}c@{\hspace{2mm}}c@{\hspace{2mm}}c@{\hspace{2mm}}c}
\includegraphics[width=0.178\textwidth]{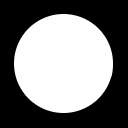} &
\includegraphics[width=0.178\textwidth]{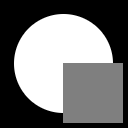} &
\includegraphics[width=0.178\textwidth]{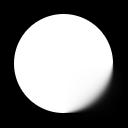} &
\includegraphics[width=0.178\textwidth]{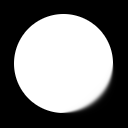} &
\includegraphics[width=0.178\textwidth]{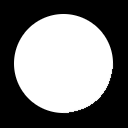} \\[1mm]
\fbox{\includegraphics[width=0.17\textwidth]{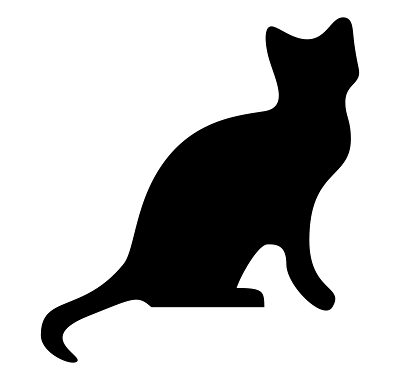}} &
\fbox{\includegraphics[width=0.17\textwidth]{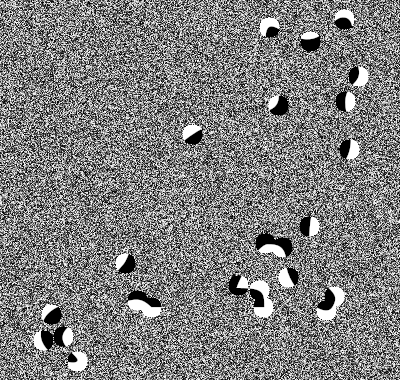}} &
\fbox{\includegraphics[width=0.17\textwidth]{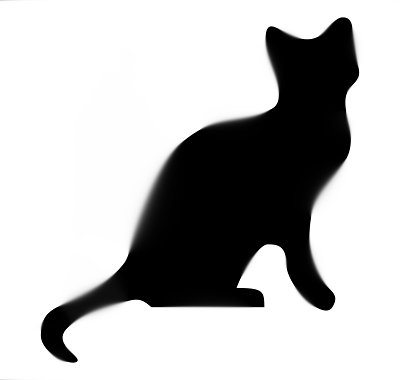}} &
\fbox{\includegraphics[width=0.17\textwidth]{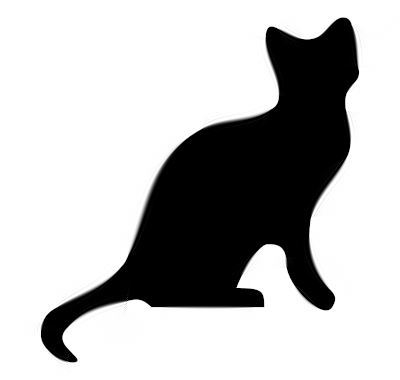}} &
\fbox{\includegraphics[width=0.17\textwidth]{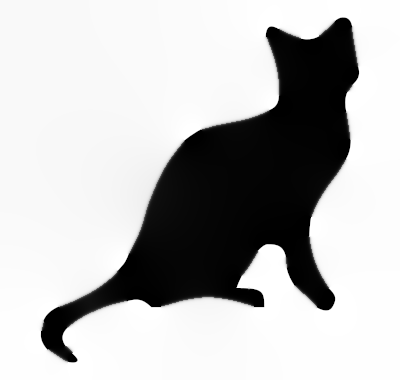}} \\
original & input & elastica & EED & DS 
\end{tabular}
\caption{Comparison of Euler's elastica, EED, and DS inpainting.
Parameters for DS inpainting:
\textbf{Top}: $\sigma = 3.2$, $\rho=3$, $\nu=3$, and $\lambda = 2$;
\textbf{Bottom:}
$\sigma = 4.2$, $\rho=4.8$, $\nu=4.5$, and $\lambda = 7$.}
\label{fig:cat}
\end{figure}

\smallskip
In Fig.~\ref{fig:cat}, we consider the shape completion of a disk and
a cat image, and compare DS inpainting with two advanced competitors: 
Euler's elastica \cite{Mu94a} and EED inpainting \cite{SPME14}.
The elastica results have been published in \cite{SAWE22}, while 
the cat data and its result for EED go back to \cite{We12}. While 
all methods accomplish their tasks, DS inpainting produces the sharpest 
edges. 

\begin{figure}
\centering
\begin{tabular}{c@{\hspace{2mm}}c@{\hspace{2mm}}c}
\includegraphics[width=0.3\textwidth]{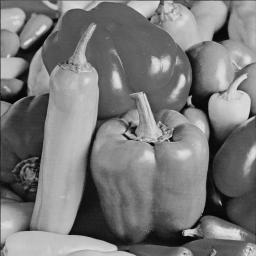} &
\includegraphics[width=0.3\textwidth]{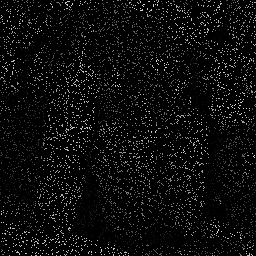}&
\includegraphics[width=0.3\textwidth]{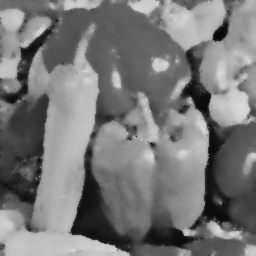}\\[1mm]
\includegraphics[width=0.3\textwidth]{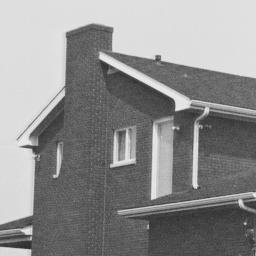} &
\includegraphics[width=0.3\textwidth]{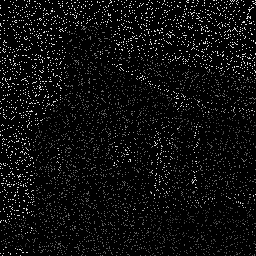}&
\includegraphics[width=0.3\textwidth]{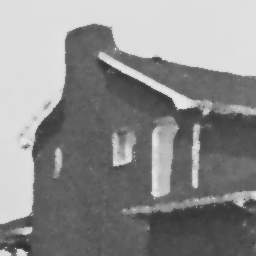}
\\
original images & inpainting data & DS inpainting\\
\end{tabular}
\caption{Inpainting of sparse greyscale images (size $256\times 256$, 
10 \% density, randomly chosen) with DS inpainting. Parameters: 
\textbf{Top}: $\sigma=2$, $\rho=1.5$, $\nu=5$, and $\lambda=3$;
\textbf{Bottom}: $\sigma=2.5$, $\rho=1.8$, $\nu=3.5$, and $\lambda=3$.}
\label{fig:grey}
\end{figure}

\smallskip
Fig.~\ref{fig:grey} shows that DS inpainting is also a powerful method for 
the reconstruction of greyscale images from sparse data. In both examples, 
the data were given by 10\% randomly selected pixels of a
$256 \times 256$ image. On a PC with an 
Intel\textsuperscript{\textcopyright} Core\texttrademark i9-11900K CPU @ 
3.50 GHz, the runtime was 
$3.32$ seconds for the {\em peppers} image 
and $3.76$ seconds for the {\em house} image. 

%
%
%
 
%
%

\section{Conclusions and Future Work}
\label{sec:conclusion}

With DS inpainting we have introduced an approach that aims
at maximal simplicity while satisfying widely accepted requirements
on inpainting methods. These requirements include the ability 
to bridge large gaps and the potential to offer sharp edges, 
a high degree of rotation invariance, and stability in terms 
of a maximum--minimum principle that prevents over- and undershoots.

\smallskip
Interestingly, this was possible by combining two ``time-honoured''
components: Homogeneous diffusion has been axiomatically introduced 
to image analysis 61 years ago \cite{Ii62}, and coherence-enhancing 
shock filters have been around for 20 years \cite{We03}. This 
demonstrates that classical methods still offer a huge potential 
that awaits being explored.

\smallskip
While most PDEs for inpainting are elliptic or parabolic, our results 
emphasise that the hyperbolic ones deserve far more attention. Hyperbolic 
PDEs are a natural concept for modelling discontinuities, and shock 
filters are a prototype for this. While their early representatives 
\cite{KB75,OR90} produced oversegmentations and were highly sensitive 
w.r.t.~to noise, more advanced variants such as coherence-enhancing 
shock filters have changed the game: Their performance reflects the 
high robustness of the structure tensor that guides them. 

\smallskip
Last but not least, the fact that DS inpainting is a second order
integrodifferential process that offers the full performance of
higher order inpainting PDEs questions the necessity of higher
order methods in practice. Without doubt, the latter ones are
algorithmically far more challenging. As a general principle in 
science, Occam's razor suggests to prefer the simplest model that 
accomplishes a desired task. Our paper adheres to this principle.


\smallskip
In our ongoing work, we are aiming at a deeper understanding of 
such integrodifferential processes, and we are investigating 
alternative applications of this promising class of methods.
  


\subsubsection{Acknowledgements.} We thank Karl Schrader for providing 
us with the images and results from his publication \cite{SAWE22}. 

\bibliographystyle{splncs04}
\bibliography{refs.bib}

\begin{thebibliography}{10}
\providecommand{\url}[1]{\texttt{#1}}
\providecommand{\urlprefix}{URL }
\providecommand{\doi}[1]{https://doi.org/#1}

\bibitem{AM94}
Alvarez, L., Mazorra, L.: Signal and image restoration using shock filters and
  anisotropic diffusion. SIAM Journal on Numerical Analysis  \textbf{31},
  590--605 (1994)

\bibitem{BEG07}
Bertozzi, A.L., Esedoglu, S., Gillette, A.: Inpainting of binary images using
  the {C}ahn–{H}illiard equation. IEEE Transactions on Image Processing
  \textbf{16}(1),  285--291 (Dec 2007)

\bibitem{BM07}
Bornemann, F., M\"arz, T.: Fast image inpainting based on coherence transport.
  Journal of Mathematical Imaging and Vision  \textbf{28}(3),  259--278 (Jul
  2007)

\bibitem{BM92}
Brockett, R.W., Maragos, P.: Evolution equations for continuous-scale
  morphology. In: Proc.~IEEE International Conference on Acoustics, Speech and
  Signal Processing. vol.~3, pp. 125--128. San Francisco, {CA} (Mar 1992)

\bibitem{BHS09}
Burger, M., He, L., Sch\"onlieb, C.: Inpainting of binary images using the
  {C}ahn–{H}illiard equation. SIAM Journal on Imaging Sciences  \textbf{2},
  1129--11671 (Nov 2009)

\bibitem{Ca88}
Carlsson, S.: Sketch based coding of grey level images. Signal Processing
  \textbf{15},  57--83 (1988)

\bibitem{CBAB97}
Charbonnier, P., Blanc-F\'eraud, L., Aubert, G., Barlaud, M.: Deterministic
  edge-preserving regularization in computed imaging. IEEE Transactions on
  Image Processing  \textbf{6}(2),  298--311 (1997)

\bibitem{Du76}
Duchon, J.: Interpolation des fonctions de deux variables suivant le principe
  de la flexion des plaques minces. RAIRO Analyse Num\'erique  \textbf{10},
  5--12 (1976)

\bibitem{DH73}
Duda, R.O., Hart, P.E.: Pattern Classification and Scene Analysis. Wiley, New
  York (1973)

\bibitem{EL99a}
Efros, A.A., Leung, T.: Texture synthesis by non-parametric sampling. In:
  Proc.~Seventh International Conference on Computer Vision. vol.~2, pp.
  1033--1038. IEEE Computer Society Press, Kerkyra, Greece (Sep 1999)

\bibitem{FG87}
F\"orstner, W., G\"ulch, E.: A fast operator for detection and precise location
  of distinct points, corners and centres of circular features. In: Proc.~ISPRS
  Intercommission Conference on Fast Processing of Photogrammetric Data. pp.
  281--305. Interlaken, Switzerland (Jun 1987)

\bibitem{Ii62}
Iijima, T.: Basic theory on normalization of pattern (in case of typical
  one-dimensional pattern). Bulletin of the Electrotechnical Laboratory
  \textbf{26},  368--388 (1962), in Japanese

\bibitem{KW22}
K\"amper, N., Weickert, J.: Domain decomposition algorithms for real-time
  homogeneous diffusion inpainting in {4K}. In: Proc.~2022 IEEE International
  Conference on Acoustics, Speech and Signal Processing. pp. 1680--1684.
  Singapore (May 2022)

\bibitem{KTZ19}
Kang, S., Tai, X.C., Zhu, W.: Survey of fast algorithms for {E}uler's
  elastica-based image segmentation. In: Kimmel, R., Tai, X.C. (eds.)
  Processing, Analyzing and Learning of Images, Shapes, and Forms: Part 2,
  Handbook of Numerical Analysis, vol.~20, pp. 533--552. Elsevier (2019)

\bibitem{KDA97}
Kornprobst, P., Deriche, R., Aubert, G.: Image coupling, restoration and
  enhancement via {PDEs}. In: Proc.~1997 IEEE International Conference on Image
  Processing. vol.~4, pp. 458--461. Washington, DC (Oct 1997)

\bibitem{KB75}
Kramer, H.P., Bruckner, J.B.: Iterations of a non-linear transformation for
  enhancement of digital images. Pattern Recognition  \textbf{7},  53--58
  (1975)

\bibitem{MHWT12}
Mainberger, M., Hoffmann, S., Weickert, J., Tang, C.H., Johannsen, D., Neumann,
  F., Doerr, B.: Optimising spatial and tonal data for homogeneous diffusion
  inpainting. In: Bruckstein, A.M., ter Haar~Romeny, B., Bronstein, A.M.,
  Bronstein, M.M. (eds.) Scale Space and Variational Methods in Computer
  Vision, Lecture Notes in Computer Science, vol.~6667, pp. 26--37. Springer,
  Berlin (2012)

\bibitem{MM98a}
Masnou, S., Morel, J.M.: Level lines based disocclusion. In: Proc.~1998 IEEE
  International Conference on Image Processing. vol.~3, pp. 259--263. Chicago,
  IL (Oct 1998)

\bibitem{Mu94a}
Mumford, D.: Elastica and computer vision. In: Bajaj, C.L. (ed.) Algebraic
  Geometry and its Applications, vol.~5681, chap.~31, pp. 491--506. Springer,
  New York (1994)

\bibitem{NR22}
Novak, A., Reini\'c, N.: Shock filter as the classifier for image inpainting
  problem using the {C}ahn--{H}illiard equation. Computers and Mathematics with
  Applications  \textbf{123},  105--114 (Oct 2022)

\bibitem{OR90}
Osher, S., Rudin, L.I.: Feature-oriented image enhancement using shock filters.
  SIAM Journal on Numerical Analysis  \textbf{27},  919--940 (1990)

\bibitem{PM90}
Perona, P., Malik, J.: Scale space and edge detection using anisotropic
  diffusion. IEEE Transactions on Pattern Analysis and Machine Intelligence
  \textbf{12},  629--639 (1990)

\bibitem{RT92}
Rouy, E., Tourin, A.: A viscosity solutions approach to shape-from-shading.
  SIAM Journal on Numerical Analysis  \textbf{29}(3),  867--884 (Jul 1992)

\bibitem{SPME14}
Schmaltz, C., Peter, P., Mainberger, M., Ebel, F., Weickert, J., Bruhn, A.:
  Understanding, optimising, and extending data compression with anisotropic
  diffusion. International Journal of Computer Vision  \textbf{108}(3),
  222--240 (Jul 2014)

\bibitem{SAWE22}
Schrader, K., Alt, T., Weickert, J., Ertel, M.: {CNN}-based {E}uler’s
  elastica inpainting with deep energy and deep image prior. In: 10th European
  Workshop on Visual Information Processing (EUVIP). Lisbon (Oct 2022)

\bibitem{SC02}
Shen, J., Chan, T.F.: Mathematical models for local non-texture inpaintings.
  SIAM Journal on Numerical Analysis  \textbf{62}(3),  1019--1043 (2002)

\bibitem{So04}
Soille, P.: Morphological Image Analysis. Springer, Berlin, second edn. (2004)

\bibitem{Ts06}
Tschumperl\'e, D.: Fast anisotropic smoothing of multi-valued images using
  curvature-preserving {PDE}'s. International Journal of Computer Vision
  \textbf{68}(1),  65--82 (Jun 2006)

\bibitem{Bo02}
{van den Boomgaard}, R.: Decomposition of the {Kuwahara--Nagao} operator in
  terms of linear smoothing and morphological sharpening. In: Talbot, H.,
  Beare, R. (eds.) Mathematical Morphology: Proc. Sixth International
  Symposium. pp. 283--292. {CSIRO} Publishing, Sydney, Australia (Apr 2002)

\bibitem{We03}
Weickert, J.: Coherence-enhancing shock filters. In: Michaelis, B., Krell, G.
  (eds.) Pattern Recognition, Lecture Notes in Computer Science, vol.~2781,
  pp.~1--8. Springer, Berlin (2003)

\bibitem{We12}
Weickert, J.: {Mathematische Bildverarbeitung mit Ideen aus der Natur}.
  Mitteilungen der DMV  \textbf{20},  80--92 (2012)

\bibitem{WW06}
Weickert, J., Welk, M.: Tensor field interpolation with {PDEs}. In: Weickert,
  J., Hagen, H. (eds.) Visualization and Processing of Tensor Fields, pp.
  315--325. Springer, Berlin (2006)

\bibitem{WW21}
Welk, M., Weickert, J.: {PDE} evolutions for {M}-smoothers in one, two, and
  three dimensions. Journal of Mathematical Imaging and Vision  \textbf{63},
  157--185 (Feb 2021)

\bibitem{WWG07}
Welk, M., Weickert, J., Gali\'c, I.: Theoretical foundations for spatially
  discrete {1-D} shock filtering. Image and Vision Computing  \textbf{25}(4),
  455--463 (2007)

\end{thebibliography}

%
%
%
%
\end{document}